\documentclass[10pt,notitlepage,a4paper,aps,prd,tightenline,preprintnumbers,nofootinbib,superscriptaddress]{revtex4-2}
\usepackage{relsize}
\usepackage[left=2cm,right=2cm,top=2.5cm,bottom=2.5cm]{geometry}
\usepackage{amsmath,amssymb,amsthm}
\usepackage{mathtools}
\usepackage{enumitem}
\usepackage{bm}
\usepackage[table,xcdraw,dvipsnames]{xcolor}
\usepackage[caption=false]{subfig}

\definecolor{NiceURL}{HTML}{A9341F}
\definecolor{NiceLinks}{HTML}{A9341F}%{524fc0}
\definecolor{NiceCite}{HTML}{0A05B5}
\definecolor{CustomBlue}{HTML}{0071BC}

\usepackage[linktocpage]{hyperref}
\hypersetup{
    colorlinks=true,
    citecolor=NiceCite,
    linkcolor=NiceLinks,
    urlcolor=NiceURL
}
\usepackage{orcidlink}

\usepackage[T1]{fontenc}
\usepackage{XCharter}

\newcommand{\N}{\mathbb{N}}

\begin{document}
\title{\Large\bf Schur elimination of brane-localised Higgs mass spectra, and the criteria for the $\epsilon\to0$/$N\to\infty$ non-commutativity}

\author{Roberto Barcel\'o$^{\orcidlink{0009-0005-8581-5499}}\,$}
\email{rbarceloaguilar@hotmail.com}
\affiliation{Independent Researcher, M\'alaga, Spain}

\author{Subhadip Mitra$^{\orcidlink{0000-0002-7107-0343}}\,$}
\email{subhadip.mitra@iiit.ac.in}
\affiliation{Center for Computational Natural Sciences and Bioinformatics, International Institute of Information Technology, Hyderabad 500~032, India}
\affiliation{Center for Quantum Science and Technology, International Institute of Information Technology, Hyderabad 500 032, India}

\begin{abstract}
\noindent
In many 5D extensions of the Standard Model, the Higgs boson is localised to a single point along the extra dimension, where it couples to bulk fermions. Computing the fermion mass spectrum then requires a boundary regulator, and removing it before or after summing the Kaluza-Klein tower can give different answers, as shown in Ref.~\cite{Barcelo:2014kha}. Because the coupling reaches any KK mode through only two overlap numbers, the whole tower can be eliminated exactly by a Schur complement; the disagreement between the two orders of limit then reduces to a single sum, whose discontinuity we prove directly. This discontinuity needs two conditions to appear: a coupling sharp enough to leave the sum conditionally convergent, and a second channel with complementary behaviour at the point of contact. Neither condition depends on having an extra dimension, or even a discrete tower to sum. The same test applies to contact interactions in ordinary scattering, the imaginary part behind the optical theorem, and seesaw, portal, and clockwork constructions. Whether the two conditions are met determines whether a similar ambiguity plays any role.
\end{abstract}
\maketitle

%======================================================================
\section{Introduction}
%======================================================================
\noindent
Models with a Higgs boson localised on a boundary (\emph{brane}) of an extra dimension, while fermions propagate in the bulk, are a standard ingredient of Randall-Sundrum constructions~\cite{Randall:1999ee,Gherghetta:2000qt} addressing the electroweak hierarchy problem and the flavour puzzle. Confining the Higgs to a single point has a well-known technical cost, the \emph{jump problem}~\cite{Csaki:2003sh,Azatov:2009na}: a Dirac delta-function source in the fermion equations of motion leaves certain profile functions ambiguous exactly at that point. The usual fix is to regularise -- shift the peak slightly off the boundary, or smooth it into a narrow bump of width $\epsilon R$ -- and remove the regulator, $\epsilon\to0$, only once the calculation is complete.

Reference~\cite{Barcelo:2014kha} showed that this regulator interacts unexpectedly with a second limit. Treating the Yukawa coupling as a mass mixing between the infinite tower of Kaluza-Klein (KK) modes gives an infinite mass matrix $M$, and the fermion mass spectrum depends on whether the KK truncation $N\to\infty$ or the regulator removal $\epsilon\to0$ is taken first.\footnote{As pointed out in Ref.~\cite{Barcelo:2014kha}, this ambiguity differs from the non-commutativity between $\epsilon\to0$ and $N_{\rm KK}^{\rm loop}\to\infty$, with $N_{\rm KK}^{\rm loop}$ the number of KK states exchanged in loop-level Higgs production and decay amplitudes~\cite{Malm:2013jia,Carena:2012fk,Casagrande:2010si,Azatov:2010pf}.} The two orders correspond, term by term, to two different but each individually consistent regularisation schemes -- Regularisation~I ($\epsilon\to0$ first) and Regularisation~II ($N\to\infty$ first) -- each matched analytically to a genuine five-dimensional calculation. Reaching this conclusion required expanding the characteristic equation of the infinite matrix $M^\dagger M$ into a hierarchy of nested sums (see their Appendix~A), eventually resummed using Lerch-transcendent and digamma asymptotics.

However, a Higgs confined to a single point along the extra dimension couples to the KK tower only through the value of each mode's wavefunction at that one point -- nothing else about the tower enters. This has the same structure as a contact interaction in ordinary scattering theory. Such interactions are always separable, reaching an arbitrarily large set of other states through a fixed, small number of overlap integrals rather than through the full detail of what they are coupled to. In the KK basis, the off-diagonal Yukawa block of the mass matrix is rank two for every truncation order and every regulator value $\epsilon$, a fact implicit in Ref.~\cite{Barcelo:2014kha} but not used there.

A finite-rank perturbation of a diagonal operator can always be removed exactly. The technique is known under several names: the Feshbach projection method in nuclear reaction theory~\cite{Feshbach1958,Feshbach1962}, L\"owdin partitioning in quantum chemistry~\cite{Lowdin1951}, and the Schur complement in linear algebra, whose determinant form is the Weinstein-Aronszajn formula~\cite{WA1,WA2,WA3}. We solve for the large or infinite sector algebraically, in terms of its own propagator, and feed the result back into the small set of states one actually cares about. In the extra-dimension context, the entire KK tower can be eliminated in closed form, collapsing what appeared as an infinite-dimensional spectral problem to a $2\times2$ determinant condition built from three scalar sums. Two of these agree whichever order the limits are taken; the third does not, and we obtain both values directly with a simple argument about a Fourier sine series. This single discontinuity reproduces the mass formulas of Regularisation~I and Regularisation~II found in Ref.~\cite{Barcelo:2014kha}, and with it the entire non-commutativity reported there is traced to one explicit, provable statement about one sum, rather than left as a property of a large expansion.

The rank-two structure follows only from the Higgs localisation, so the same reduction extends to a warped background once the flat KK profiles are replaced by the corresponding Bessel functions, and to the full three-generation flavour structure left open in Ref.~\cite{Barcelo:2014kha}. More generally, the same separability arises whenever a small set of states reaches a large or infinite tower through only a handful of overlap numbers, as it is the case for seesaw messengers, for portal couplings to a hidden sector, and for deconstructed extra dimensions, and the method presented here applies to each of these without modification.

The paper is organised as follows. Section~\ref{sec:edmodel} recalls the model of Ref.~\cite{Barcelo:2014kha} and shows the rank-two structure of its Yukawa coupling. Section~\ref{sec:KKelimination} eliminates the KK tower exactly, reducing the problem to a $2\times2$ determinant condition. Section~\ref{sec:sums} evaluates the resulting self-energies, isolates the single discontinuous limit behind the two mass spectra of Ref.~\cite{Barcelo:2014kha}, and extends the reduction to a warped background, to the full flavour structure, and to a smoothly regularised Higgs profile. Section~\ref{sec:generalisation} traces the same mechanism into ordinary quantum scattering, the optical theorem, and other constructions built on a coupling to a large sector through a handful of overlap numbers. We conclude in Section~\ref{sec:conclusions}.

%======================================================================
\section{The model, and why its brane coupling is separable}
\label{sec:edmodel}
%======================================================================
\noindent
We use the conventions of Ref.~\cite{Barcelo:2014kha}: a flat extra dimension with coordinate $y\in[0,\pi R]$, bulk down-type quark fields $Q,D$ (doublet and singlet of $SU(2)_L$), and a Higgs doublet $H$ confined to the brane at $y=\pi R$, with action
\begin{align}
    S_{\rm fermion}=\int d^4x\,dy&\ \Big[\tfrac{i}{2}\big(\bar Q\Gamma^M\partial_M Q-\partial_M\bar Q\Gamma^M Q+\{Q\leftrightarrow D\}\big)\nonumber\\
    &\  -\delta(y-\pi R)\big(Y_5\,\bar Q_LHD_R+Y_5'\,\bar Q_RHD_L+{\rm H.c.}\big)\Big].    
\end{align}
Expanding the 5D fields in the free (Yukawa-less) KK basis, we get
\begin{align}
    &Q_L=\sum_n q_L^n(y)Q_L^n(x), &&Q_R=\sum_n q_R^n(y)Q_R^n(x),\nonumber\\
    &D_L=\sum_n d_L^n(y)D_L^n(x), &&D_R=\sum_n d_R^n(y)D_R^n(x),    
\end{align}
with the normalised zero-order profiles with $(--)$/$(++)$ boundary conditions given as
\begin{align}
    q_L^n(y)=d_R^n(y)&=\sqrt{\tfrac{2}{\pi R}}\cos\!\Big(\tfrac{ny}{R}\Big)\ (n>0),\qquad
    q_L^0=d_R^0=\sqrt{\tfrac1{\pi R}},\label{eq:profQL}\\
    -q_R^n(y)=d_L^n(y)&=\sqrt{\tfrac{2}{\pi R}}\sin\!\Big(\tfrac{ny}{R}\Big)\ (n>0),\qquad q_R^0=d_L^0=0,
\label{eq:profQR}
\end{align}
and the KK masses as $M_n\equiv M_{qn}=M_{dn}=n/R$. 

In the combined basis $\Psi_L^t=(Q_L^0,D_L^0,Q_L^1,D_L^1,\dots)$, $\Psi_R^t=(Q_R^0,D_R^0,Q_R^1,D_R^1,\dots)$, the effective 4D Lagrangian is 
\begin{equation}
    \mathcal L_{\rm mass}=-\bar\Psi_L\cdot[M]\cdot\Psi_R+{\rm H.c.},    
\end{equation}
with the infinite matrix,
\begin{align*}
    &M_{Q_L^i,Q_R^j}=M_{qi}\,\delta_{ij},\quad\qquad
    M_{D_L^i,D_R^j}=M_{di}\,\delta_{ij},
    &&M_{Q_L^i,D_R^j}=\alpha_{ij},\qquad\quad
    M_{D_L^i,Q_R^j}=\beta_{ij},\nonumber\\    
    &\alpha_{ij}=Y_5\int_0^{\pi R}\!dy\,\delta(y-\pi R)\,\tfrac{v}{\sqrt2}\,q_L^i(y)\,d_R^j(y),
    &&\beta_{ij}=Y_5'\int_0^{\pi R}\!dy\,\delta(y-\pi R)\,\tfrac{v}{\sqrt2}\,d_L^i(y)\,q_R^j(y),
\end{align*}
i.e., 
\begin{equation*}
    [M]=
    \begin{pmatrix}
    M_{0} & \alpha_{00} & 0 & \alpha_{01} & 0 & \alpha_{02} & \cdots\\
    \beta_{00} & M_{0} & \beta_{01} & 0 & \beta_{02} & 0 & \cdots\\
    0 & \alpha_{10} & M_{1} & \alpha_{11} & 0 & \alpha_{12} & \cdots\\
    \beta_{10} & 0 & \beta_{11} & M_{1} & \beta_{12} & 0 & \cdots\\
    0 & \alpha_{20} & 0 & \alpha_{21} & M_{2} & \alpha_{22} & \cdots\\
    \beta_{20} & 0 & \beta_{21} & 0 & \beta_{22} & M_{2} & \cdots\\
    \vdots & \vdots & \vdots & \vdots & \vdots & \vdots & \ddots
    \end{pmatrix}.
\end{equation*}

%======================================================================
\subsection{The brane coupling is separable}
\noindent
The delta function in the off-diagonal terms samples the wavefunctions exactly at one point; irrespective of what the KK profile of mode $i$ looks like along $y$, the Higgs only sees its value at the brane. The coupling between two KK modes can therefore depend on $i$ and $j$ only through the product of two single-mode numbers, one for each side of the vertex -- it cannot depend on $i$ and $j$ jointly in any other way, because the brane never has access to more than one point of either profile at a time.

The KK masses sit only on the diagonal of $M$, whereas the Yukawa coupling sits entirely in the off-diagonal terms. To make this split explicit, we arrange the basis by particle type instead of by KK level, $(Q^0,Q^1,\dots)\oplus(D^0,D^1,\dots)$. This reshuffle, applied identically on the $L$ and $R$ sides, leaves every eigenvalue and singular value of $M$ unchanged. Writing $M=\hat D+V$ in this basis, $\hat D$ is the free KK mass operator,
\begin{equation}
    \hat D=\begin{pmatrix}D_0&0\\0&D_0\end{pmatrix},
    \qquad D_0={\rm diag}(M_0,M_1,M_2,\dots),
\end{equation}
and $V$ collects the two off-diagonal blocks $[\alpha]$ and $[\beta]$, the only piece of $M$ that the Higgs is responsible for:
\begin{equation}
    V=\begin{pmatrix}0&[\alpha]\\ [\beta]&0\end{pmatrix}.
\end{equation}
The diagonal matrix, $D_0$, mixes each left-handed mode only with its own right-handed partner of the same type, $Q_L^n$ with $Q_R^n$ and $D_L^n$ with $D_R^n$, while $V$ carries only the two Yukawa-induced cross terms: $Q_L^i$ mixes with $D_R^j$ through $\alpha_{ij}$, and $D_L^i$ mixes with $Q_R^j$ through $\beta_{ij}$.

Ref.~\cite{Barcelo:2014kha} regularises the jump problem~\cite{Csaki:2003sh} in two equivalent ways; we use the one that keeps this structure exact at any finite $\epsilon$, shifting the delta peak away from the boundary, $\delta(y-\pi R)\to\delta(y-y_\star)$, with $y_\star\equiv(\pi-\epsilon)R$. The matrix elements reduce to the profile functions evaluated at $y_\star$:
\begin{equation}
    \alpha_{ij}=X\,q_L^i(y_\star)\,d_R^j(y_\star),\qquad
    \beta_{ij}=X'\,d_L^i(y_\star)\,q_R^j(y_\star),\qquad
    X\equiv\frac{vY_5}{\sqrt2},\quad X'\equiv\frac{vY_5'}{\sqrt2}.
\label{eq:factorized}
\end{equation}
Defining, for $n\geqslant0$,
\begin{equation}
    f_n\equiv q_L^n(y_\star)=d_R^n(y_\star),\qquad k_n\equiv q_R^n(y_\star)=-d_L^n(y_\star),
    \label{eq:fkdef}
\end{equation}
Eqs.~\eqref{eq:profQL} and~\eqref{eq:profQR} give
\begin{align}
    &f_0=\sqrt{\frac1{\pi R}},&& f_n=\sqrt{\frac2{\pi R}}(-1)^n\cos(n\epsilon)\ (n\geqslant1),   \nonumber\\
    &k_0=0, &&k_n=\sqrt{\frac2{\pi R}}(-1)^n\sin(n\epsilon)\ (n\geqslant1), \label{eq:kndef}
\end{align}
so that Eq.~\eqref{eq:factorized} becomes, in operator form,
\begin{equation}
    V=X\,\big|f\big\rangle_{L_Q}\!\big\langle f\big|_{R_D}\ -\ X'\,\big|k\big\rangle_{L_D}\!\big\langle k\big|_{R_Q}\ ,\label{eq:Vrank2}
\end{equation}
using the same vector $f$ on both sides since $q_L^n(y_\star)=d_R^n(y_\star)$ by Eq.~\eqref{eq:fkdef}, and likewise for $k$ up to the sign $d_L^n(y_\star)=-k_n$. Whatever the truncation order $N$ and whatever the value of the regulator $\epsilon$, $V$ is built from only these two vectors, $f,k\in\ell^2(\N_0)$: every KK mode couples to the brane through nothing more than its own entry in $f$ or $k$.

The same argument, however, does not survive smoothing the peak into a square profile of width $\epsilon R$. There the coupling involves an integral of the profiles over an interval, not their value at a point, and at finite $\epsilon$ the brane genuinely samples more than one point of each mode -- the separable structure only reappears in the strict $\epsilon\to0$ limit. We return to this point in Sec.~\ref{sec:sums}.

%======================================================================
\section{Integrating out the Kaluza-Klein tower}
\label{sec:KKelimination}
%======================================================================
\subsection{Physical set-up}
\noindent
The physical fermion masses are the singular values of $M$, i.e., they are nonzero $m$ for which there exist $(p,q)\in L_Q\oplus L_D$ and $(y,x)\in R_Q\oplus R_D$, not both zero, such that
\begin{equation}
M(y,x)=m\,(p,q),\qquad M^{T}(p,q)=m\,(y,x).
\label{eq:sv}
\end{equation}
Equivalently, $m^2$ is an eigenvalue of $M^TM$, the matrix diagonalised in Ref.~\cite{Barcelo:2014kha}. However, working directly with the pair Eq.~\eqref{eq:sv} gives us slightly more information, as we see below. Writing $M=\hat D+V$ with $V$ as in Eq.~\eqref{eq:Vrank2}, and denoting by $\hat D$ the common diagonal operator $D_0={\rm diag}(M_n)$ acting on either the $Q$ or $D$ sector, Eq.~\eqref{eq:sv} becomes the coupled linear system
\begin{align}
\hat Dy+Xf\,(f\!\cdot\! x)&=mp, &\hat Dx-X'k\,(k\!\cdot\! y)&=mq,\label{eq:sys1}\\
\hat Dp-X'k\,(k\!\cdot\! q)&=my, &\hat Dq+Xf\,(f\!\cdot\! p)&=mx.\label{eq:sys2}
\end{align}
The KK tower enters this system only through four numbers, the scalar collective coordinates:
\begin{equation}
a\equiv f\!\cdot\! x,\qquad b\equiv k\!\cdot\! y,\qquad c\equiv f\!\cdot\! p,\qquad d\equiv k\!\cdot\! q.
\label{eq:abcd}
\end{equation}

%======================================================================
\noindent
\subsection{Eliminating the tower}
\noindent
Let us now take any $m\neq0$ and work at a finite truncation order $N$, so every inversion below is algebra on finite vectors; we take $N\to\infty$ only afterwards, in the resulting closed-form sums. From Eq.~\eqref{eq:sys1}, $\hat Dy=mp-Xaf$; combined with
Eq.~\eqref{eq:sys2}, $\hat Dp=my-X'dk$, substitution gives
\begin{align}
    &(\hat D^2-m^2)p=X'd\,\hat Dk-mXa\,f\nonumber \\ 
    \Rightarrow\quad &p=X'd\,(\hat D^2-m^2)^{-1}\hat Dk-mXa\,(\hat D^2-m^2)^{-1}f,    
\end{align}
and, by the symmetry $(y,p,a,f,X)\leftrightarrow(x,q,b,k,-X')$ of
Eqs.~\eqref{eq:sys1}-\eqref{eq:sys2}, we get
\begin{align*}
    &q=mX'b\,(\hat D^2-m^2)^{-1}k-Xc\,(\hat D^2-m^2)^{-1}\hat Df,\\ &x=X'b\,(\hat D^2-m^2)^{-1}\hat Dk-mXc\,(\hat D^2-m^2)^{-1}f,\\
    &y=mX'd\,(\hat D^2-m^2)^{-1}k-Xa\,(\hat D^2-m^2)^{-1}\hat Df.
\end{align*}
Thus, we have expressed the solutions of the coupled equations in terms of $(\hat D^2-m^2)^{-1}$ and the four unknown scalars $(a,b,c,d)$.

Solving these equations introduces three self-energy terms:
\begin{align*}
    &F(m)\equiv f\cdot(\hat D^2-m^2)^{-1}f=\sum_{n}\frac{f_n^2}{M_n^2-m^2},\qquad
K(m)\equiv k\cdot(\hat D^2-m^2)^{-1}k=\sum_n\frac{k_n^2}{M_n^2-m^2},
\label{eq:FKPdef}    \\
&P(m)\equiv f\cdot(\hat D^2-m^2)^{-1}\hat D\,k=\sum_n\frac{f_nk_nM_n}{M_n^2-m^2}=k\cdot(\hat D^2-m^2)^{-1}\hat D\,f,
\end{align*}
where $F$ and $K$ are the self-energy contributions from virtual exchange of the whole KK tower for $f$ and $k$, and $P$ is the corresponding mixed-chirality term. With these, contracting $p,q,x,y$ with $f$ or $k$ gives us four algebraic equations for $a,b,c,d$:
\begin{equation}
\begin{aligned}
a&=X'P\,b-mXF\,c, & c&=X'P\,d-mXF\,a,\\
b&=mX'K\,d-XP\,a, & d&=mX'K\,b-XP\,c.
\end{aligned}
\label{eq:closed4}
\end{equation}
Since the system is invariant under $(a,b)\leftrightarrow(c,d)$, we set $s=a+c,\ t=a-c,\ u=b+d,\ w=b-d$ to split the system into a symmetric sector $(s,u)$ and an antisymmetric sector $(t,w)$:
\begin{align}
(1+mXF)\,s-X'P\,u&=0, & XP\,s+(1-mX'K)\,u&=0,\label{eq:sym}\\
(1-mXF)\,t-X'P\,w&=0, & XP\,t+(1+mX'K)\,w&=0.\label{eq:antisym}
\end{align}
For non-trivial physical modes to exist, the determinant of both systems must vanish, giving us
\begin{align}
    \big(1+mXF\big)\big(1-mX'K\big)+XX'P^2=0&\qquad\text{(symmetric sector)},\label{eq:CEsym}    \\
    \big(1-mXF\big)\big(1+mX'K\big)+XX'P^2=0&\qquad\text{(antisymmetric sector).}\label{eq:CEantisym}
\end{align}
Equations~\eqref{eq:CEsym} and~\eqref{eq:CEantisym} are exchanged under $m\to-m$, since $F$ and $K$ are even and $P$ is odd in $m$; they encode the same physical (unsigned) mass spectrum, and their product is a quartic in $\cot(\pi mR)$, matching the structure of Ref.~\cite{Barcelo:2014kha}.

Now, a finite set of distinguished states coupled to a large or infinite background only through a few numbers -- e.g., here, the four coordinates $a,b,c,d$ -- can always have that background solved for algebraically and folded back in, leaving a finite problem however large the background is. This is the Schur complement in matrix analysis, whose determinant form is the well-known \emph{matrix determinant lemma},
\begin{equation}
\det\!\big(A+UW^T\big)=\det(A)\,\det\!\big(I_r+W^TA^{-1}U\big),
\end{equation}
where $A$ is an $n\times n$ invertible matrix, and $U,W$ are $n\times r$ matrices with $r\leqslant n$. We prove this identity, and its restatement as the elimination of the linear system above, in Appendix~\ref{app:mdl}.

%======================================================================
\section{A coupling that vanishes mode by mode, but not in total}
\label{sec:sums}
%======================================================================
\noindent
Two of the three self-energy terms drop out immediately. The $F$ term is built from $f_n$, the profile of the right-chirality modes at the brane, and these do not vanish as $\epsilon\to0$; correspondingly, $F$ settles smoothly onto a single value, whichever order the tower is summed and the regulator removed. Using $f_n^2=(2/\pi R)\cos^2(n\epsilon)$ ($n\geqslant1$), $f_0^2=1/\pi R$, and writing $x\equiv mR$, we get
\begin{equation}
    F(m)=-\frac1{\pi Rm^2}+\frac{2R}\pi\sum_{n=1}^\infty\frac{\cos^2(n\epsilon)}{n^2-x^2}.
\end{equation}
Since $\cos^2(n\epsilon)$ is bounded and $1/(n^2-x^2)$ is absolutely summable away from the integers, $\epsilon\to0$ can be taken inside the sum regardless of when the tower is truncated, leaving the classical Mittag-Leffler expansion of the cotangent, $\pi x\,\cot(\pi x)=1+\sum_{n=1}^\infty 2x^2/(x^2-n^2)$. Hence, we get
\begin{equation}
    F(m)=-\frac{\cot(\pi mR)}{m}.
    \label{eq:Fresult}
\end{equation}
The $K$ term, built the same way from $k_n$, is even smaller: $k_n\to0$ as $\epsilon\to0$ for every $n$, since the left-handed profiles vanish at the brane, and the same dominated-convergence argument gives simply
\begin{equation}
    K(m)=0,\label{eq:Kresult}
\end{equation}
whichever order the two limits are taken.  

The $P$ term is different, as it is built from the cross-chirality product $f_nk_n$ -- one profile that survives at the brane, one that does not. If we take $\epsilon\to0$ first, every term in the series vanishes, making $P=0$. If instead we sum the series first, we get an $\epsilon$-independent term that does not vanish. The reason is a rescaling, not a cancellation: writing $u=n\epsilon$, each term in the sum is proportional to $\epsilon$, but the spacing between successive values of $u$ is also $\epsilon$, so the sum over $n$ is a Riemann sum, of mesh size $\epsilon$, for an integral over $u$. As $\epsilon\to0$ the mesh refines and the sum converges to that integral, which is finite and independent of $\epsilon$ by construction -- the vanishing of each term and the growing density of terms cancel exactly, term for term. This is the same mechanism behind the overshoot in a truncated Fourier series evaluated at a jump, where a sum of individually shrinking terms that reorganises itself, in the appropriate rescaled variable, into a fixed continuum integral.

Using $f_nk_n=\frac1{\pi R}\sin(2n\epsilon)$ and $M_n/(M_n^2-m^2)=\frac12\big[\frac1{n-x}+\frac1{n+x}\big]R$, we can write
\begin{equation}
    P(m)=\frac1{2\pi}\left(\sum_{n=1}^\infty\frac{\sin(2n\epsilon)}{n-x}+\sum_{n=1}^\infty\frac{\sin(2n\epsilon)}{n+x}\right), \label{eq:Pdef}
\end{equation}
where the $n=0$ term vanishes since $k_0=0$. If the sum is carried out over the full tower first, each sine series gives $\pi/2$ (we prove this in Appendix~\ref{app:sine}, where the same rescaling appears as the classical Dirichlet integral,
$\int_0^\infty(\sin u/u)\,du=\pi/2$). Hence, the two orders of limit disagree:
\begin{equation}
    \lim_{\epsilon\to0^+}\lim_{N\to\infty}P=\frac12\ \neq\ 0=\lim_{N\to\infty}\lim_{\epsilon\to0}P.\label{eq:noncommute}
\end{equation}

The equation governing a single KK profile is the textbook particle-in-a-box problem. A particle confined to an interval of width $L$ with a point interaction $\lambda\delta(x-x_0)$ inside it obeys $c_n(E_n-E)=-\lambda\,\psi(x_0)\phi_n(x_0)$, which closes into $1+\lambda\,G(E,x_0)=0$ with $G=\sum_n\phi_n(x_0)^2/(E_n-E)$ -- the same object as $F$, reached by the same elimination, continuous in the location of the point interaction for the same reason $F$ is continuous in $\epsilon$. The extra dimension here is two such boxes, one per chirality, and the discontinuity above could not come from either alone: it needs the cross term $P$, built from two profiles with opposite behaviour at the brane, which a single channel cannot produce.

%======================================================================
\subsection{The two mass spectra from one number}
\noindent
Feeding $F$, $K$, and the two values of $P$ back into Eq.~\eqref{eq:CEsym} turns this single discontinuity into two different mass formulas. Setting $\epsilon\to0$ before summing the tower gives $P=K=0$, and Eq.~\eqref{eq:CEsym} collapses to $1+mXF=0$, i.e.,
\begin{equation}
    \tan(\pi mR)=\frac{vY_5}{\sqrt2}.\label{eq:regI}
\end{equation}
This is the mass spectrum obtained when the cross-chirality coupling is switched off before the tower is allowed to respond collectively. Summing the tower first, Eq.~\eqref{eq:CEsym} instead gives 
\begin{equation}
    \tan(\pi mR)=\frac{X}{1+XX'/4}=\frac{4\sqrt2\,vY_5}{8+v^2Y_5Y_5'}.\label{eq:regII}
\end{equation}
Here, the cross-chirality tower has already built up its coherent response before the regulator is removed, and that response survives as the extra term. We have thus independently arrived at both mass spectra of Ref.~\cite{Barcelo:2014kha}. Our derivation changes only the method, not the physical conclusion; the recommendation to build the theory with the complete tower before applying any cutoff is unaffected by how we eliminate the tower.

%======================================================================
\subsection{Extending the reduction: warped backgrounds and flavour generalisation}
\noindent
A warped background changes the profiles to Bessel functions and the KK masses to the corresponding Bessel zeros, but the coupling remains a value at one point, so $V$ stays exactly rank two, for any truncation and any $\epsilon$. The elimination in Sec.~\ref{sec:KKelimination} is linear algebra on a rank-two perturbation of a diagonal operator, and uses only that $\hat D$ is diagonal, never its spacing. The reduction to a $2\times2$ determinant condition therefore carries over directly, with $\hat D$ the warped KK mass operator in place of $\mathrm{diag}(n/R)$. The discontinuity survives for the same structural reason. The mechanism in Sec.~\ref{sec:sums} turned on $n\epsilon$ becoming a continuous variable as $\epsilon\to0$, converting the tail of the sum into a Riemann integral over an equally spaced tower. Bessel zeros are asymptotically equally spaced at large $n$, precisely the regime that the Riemann sum probes, so the same jump is expected in the warped case. 

The flavour generalisation changes even less. Promoting $Y_5,Y_5'$ to $3\times3$ matrices in generation space leaves the KK profiles $f_n,k_n$ untouched, since they carry no generation index in this model, so $\alpha,\beta$ stay separable in $i,j$ and $V$ becomes rank six (two KK-profile vectors combined with two $3\times3$ matrices in place of two numbers). The $F$, $K$, and $P$ terms depend only on the KK profiles and are unchanged; $X$ and $X'$ become matrices, the collective coordinates $a,b,c,d$ become vectors in generation space, and the elimination in Sec.~\ref{sec:KKelimination} closes the system into a determinant condition built from matrices rather than scalars. Reading off physical masses and mixing angles from that condition means diagonalising it, a separate calculation, but eliminating the tower needs no new argument.

%======================================================================
\subsection{Extended Higgs profiles}
\noindent
Separability fails for the smoothed square profile of width $\epsilon R$ when the full tower is kept. Ref.~\cite{Barcelo:2014kha} shows matrix elements for this profile: $\alpha_{ij}\propto\sin[(i+j)(\pi-\epsilon)]/(i+j)+\sin[(i-j)(\pi-\epsilon)]/(i-j)$, which depend on $i,j$ only through $n=i\pm j$, and reduce to the kernel $\sin(n\epsilon)/n$. This kernel is roughly constant for every $n\lesssim1/\epsilon$: smoothing the peak correlates a growing band of KK levels as $\epsilon\to0$, rather than confining the coupling to a fixed, small number of them. Within a tower truncated at $N$, this band covers the whole relevant range whenever $N\epsilon\ll1$, and a kernel constant across all its entries is rank one -- so the low-rank picture survives as long as the smoothing scale stays finer than the highest KK level kept. This can always be arranged at fixed, finite $N$ by shrinking $\epsilon$ first, and more generally along any joint limit with $\epsilon N\to0$, the regulator shrinking faster than new levels are admitted; it cannot be arranged for the full tower at any fixed $\epsilon>0$. The same asymmetry between the two orders of limit that produced the discontinuity above reappears here as a condition on the relative rate of the two limits, rather than as a discontinuity within one formula.

The problem traces back to the sharp edges of the square profile. A profile of finite width and sharp edges has matrix elements that decay only as a power law in $n$, $\sin(n\epsilon)/n\sim1/n$ for $n\epsilon\gg1$, which is what forces the $\epsilon N\to0$ condition above. A profile that vanishes smoothly at the edges of its support, together with all its derivatives -- a Gaussian bump in place of a square one, for instance -- has matrix elements that decay faster than any power of $n$, so its departure from the exact rank-two structure is negligible at any fixed $\epsilon>0$, for the entire tower, with no relative rate between $\epsilon$ and $N$ required. This can be checked directly. Replacing the square bump by a Gaussian of width $\epsilon R$ centred at $y_\star$, the same overlap integral becomes a Gaussian transform
rather than a Dirichlet kernel,
\begin{equation}
    \alpha_{ij}\propto\cos\!\Big[\frac{(i-j)y_\star}{R}\Big]e^{-(i-j)^2\epsilon^2/2}+\cos\!\Big[\frac{(i+j)y_\star}{R}\Big]e^{-(i+j)^2\epsilon^2/2},
\end{equation}
up to corrections suppressed by $e^{-O(1/\epsilon^2)}$ from extending the integral over all $y$. This is the same $n=i\pm j$ structure as the square profile, but the kernel now decays as $e^{-n^2\epsilon^2/2}$ rather than $1/n$: for $n\lesssim1/\epsilon$ it is close to constant, just as before, but for $n\gtrsim1/\epsilon$ it essentially drops to zero immediately. The matrix elements decay as $e^{-n^2\epsilon^2/2}$ rather than $1/n$, for any fixed $\epsilon>0$ and the entire tower, with no relative rate between $\epsilon$ and $N$ needed, unlike the square profile (see Appendix~\ref{app:numerics}). The departure from rank two is therefore suppressed exponentially, and at the level of precision any physical calculation works to, $V$ is rank two.

%======================================================================
\section{Two conditions, and where they hold}
\label{sec:generalisation}
%======================================================================
\noindent
There are two criteria for this jump to show up, both already visible in what came before. A single channel is not enough. We saw that a single profile could only produce the ordinary particle-in-a-box condition, continuous regardless of order; the cross term $P$, built from two profiles with complementary behaviour at the brane, is what is needed. The overlap must also be localised enough that this cross term is only conditionally, not absolutely, convergent. Neither criterion alone produces the discontinuity. In the extra-dimension model, the discontinuity comes from a conditionally convergent series, but it need not be about a discrete series at all. What pushes $P$ to the edge of absolute convergence is that the coupling reaches the tower through a value at one point, and by the uncertainty principle a coupling localised in one variable has an undamped tail in the conjugate variable. The same mechanism appears wherever a response function is built by integrating over a continuum coupled through a point.

The contact interaction in ordinary quantum mechanics offers a clear illustration. For a zero-range potential $V(\mathbf r)=g\,\delta^3(\mathbf r)$, the bound-state condition, obtained by the same elimination as Sec.~\ref{sec:KKelimination} with the tower replaced by a momentum integral, is
\begin{equation}
    1=g\int\!\frac{d^3k}{(2\pi)^3}\,\frac{1}{E-k^2/2m}.\label{eq:contact}
\end{equation}
The integral diverges linearly with the momentum cutoff $\Lambda$,
\begin{equation}
    \int_0^\Lambda\!\frac{4\pi k^2\,dk}{(2\pi)^3}\,\frac{2m}{-k^2}=-\frac{m\Lambda}{\pi^2}+O(1),
\end{equation}
so $g$ cannot be held fixed as $\Lambda\to\infty$. The low-energy scattering amplitude off this potential is fixed by a single physical number, the scattering length $a$, and demanding that it stay finite as $\Lambda\to\infty$ forces
\begin{equation}
    \frac1{g(\Lambda)}+\frac{m\Lambda}{\pi^2}=\frac{m}{2\pi a}.
\end{equation}
Unlike $P$, which varies between $0$ and $1/2$ depending on the order of the limits, here the integral diverges as $\Lambda\to\infty$. Keeping $a$ finite then requires $g(\Lambda)$ to change with the cutoff rather than sit at some fixed value. Whether one removes the cutoff first and reads off $a$ from the low-energy scattering data, or holds $g(\Lambda)$ fixed and takes $\Lambda\to\infty$ only at the end, is the same question asked of $\epsilon$ and $N$ here, now forced on the theory rather than appearing only for a particular sum.

The same regulator reappears once the contact interaction above is put on shell. If $E$ in Eq.~\eqref{eq:contact} sits on the positive real axis, the propagator $1/(E-k^2/2m)$ develops a genuine pole inside the integration range, and the integral is undefined until the pole is given a prescription -- the same kind of regulator as before, $E\to E\pm i\epsilon$. Away from the pole, the integral is absolutely convergent, but integrated through $k^2/2m=E$, it diverges from either side, and we have to consider it via its symmetric, principal-value limit. This is the continuum analogue of a sum that is only conditionally convergent, as $P$ was. The propagator splits into two pieces with different behaviour at the singularity, as $F$ and $P$ differed at the brane:
\begin{equation}
\lim_{\epsilon\to0^+}\frac1{x\mp i\epsilon}=\mathrm P\frac1x\pm i\pi\delta(x)
\end{equation}
separates a principal-value part, which survives $\epsilon\to0$ smoothly regardless of order, from a term that is present only if the pole is resolved before $\epsilon\to0$ is taken -- the same asymmetry that gave $P=1/2$ rather than $P=0$. What was a jump between two real numbers for a discrete tower becomes, on a continuum with a genuine on-shell state to reach, a jump between zero and a nonzero imaginary part -- a decay width, or the discontinuity of an amplitude across its cut. This is why the optical theorem, $\mathrm{Im}\,T_{ii}=\tfrac12\sum_f|T_{fi}|^2$, belongs to the same family: the on-shell channel supplies the second, complementary piece that $P$ needed and $F$ never had.

Similar questions can arise in model-building whenever a coupling into a large or infinite sector is built from a small number of overlap numbers rather than a generic matrix. However, unless both criteria are satisfied, the discontinuity does not appear. The case of bulk right-handed neutrinos~\cite{Dienes:1998sb,Arkani-Hamed:1998wuz} offers an illustration. A right-handed neutrino spread over a large extra dimension, with a Dirac Yukawa localised at $y_\star$ exactly as for the quarks in Sec.~\ref{sec:edmodel}, but paired with a Majorana mass for each KK level rather than a second Dirac partner, couples to KK mode $n$ through a single overlap number $f_n$. Diagonalising this seesaw-type mass matrix level by level gives
\begin{equation}
m_\nu=-v^2\sum_n\frac{f_n^2}{M_n},
\label{eq:seesaw}
\end{equation}
the ordinary seesaw formula summed over the full tower rather than truncated at the lowest level. The sum is built from a single overlap number the same way $F$ is, and single-channel sums of this kind are absolutely convergent. Hence, the construction carries none of $P$'s ambiguity. This separable structure already appears in current model-building. Warped vectorlike-quark studies build their effective Yukawa couplings from the same brane-profile products~\cite{Gopalakrishna:2013hua}, though truncating the tower by hand bypasses the order-of-limits question in most phenomenological studies. The same reasoning applies to other single-operator couplings into a tower: a Higgs or kinetic-mixing portal to a hidden sector, and clockwork constructions~\cite{Choi:2015fiu,Giudice:2016yja}, where the visible sector couples to a chain at one site, are both single channels by default, and acquire $P$'s ambiguity only if a second, oppositely-localised coupling is deliberately added.

%======================================================================
\section{Conclusions}
\label{sec:conclusions}
%======================================================================
\noindent
A brane-localised Higgs reaches the Kaluza-Klein tower through a value at one point, so its Yukawa coupling is exactly rank two, whatever the truncation and whatever the regulator. This lets the whole tower be eliminated in closed form by an elementary Schur complement, replacing the combinatorial expansion of Ref.~\cite{Barcelo:2014kha} with a $2\times2$ determinant built from three self-energies. Two of these are absolutely convergent and settle the matter regardless of order; the third, $P$, is only conditionally convergent, and summing the tower before or after removing the regulator gives $1/2$ or $0$. This number reproduces both mass spectra of Ref.~\cite{Barcelo:2014kha} and accounts for the non-commutativity reported there.

The reduction survives once the flat extra dimension is warped, once the coupling carries three generations, and once the Higgs profile is smoothed, provided it vanishes, together with all its derivatives, at its edges; it fails for a sharply-edged profile, for the reason given in Sec.~\ref{sec:sums}. The discontinuity itself needs two conditions: a second channel with complementary behaviour at the point of contact, and a coupling sharp enough to leave the resulting sum conditionally rather than absolutely convergent. Neither is peculiar to an extra dimension. The same pair governs a contact interaction in ordinary scattering, where it forces the coupling to run with a momentum cutoff, and the same locality, carried into momentum space, is what the optical theorem turns into an imaginary part. Bulk neutrino seesaws, portal couplings, and clockwork constructions meet one condition by default and the other only if deliberately arranged, which is why they are usually safe from it.

%======================================================================
\appendix
\section*{Appendix}
\section{The matrix determinant lemma}
\label{app:mdl}
%======================================================================
\noindent
Let $A$ be an invertible $n\times n$ matrix and let $U,W$ be $n\times r$ matrices. Then the lemma says
\begin{equation}
\det\!\big(A+UW^T\big)=\det(A)\,\det\!\big(I_r+W^TA^{-1}U\big).
\end{equation}
Let us consider the $(n+r)\times(n+r)$ block matrix $\begin{pmatrix}A&-U\\ W^T&I_r\end{pmatrix}$. Eliminating the lower-left block via the Schur complement on the upper-left invertible block $A$ gives
\begin{equation*}
\det\begin{pmatrix}A&-U\\ W^T&I_r\end{pmatrix}=\det(A)\det\!\big(I_r+W^TA^{-1}U\big).
\end{equation*}
Eliminating the upper-right block instead via the Schur complement on the lower-right invertible block $I_r$ gives
\begin{equation*}
\det\begin{pmatrix}A&-U\\ W^T&I_r\end{pmatrix}=\det(I_r)\det\!\big(A+UW^T\big)=\det(A+UW^T).
\end{equation*}
Equating the two expressions proves the lemma. This is the determinant form of the identity; the corresponding statement for the inverse itself is the Sherman-Morrison-Woodbury formula~\cite{Sherman:1950,Woodbury:1950}, of which the rank-one case is the classical Sherman-Morrison update. The same fact, stated for a trace-class rather than finite-rank perturbation of a self-adjoint operator, underlies Krein's spectral shift function~\cite{Krein:1953}, which quantifies how the whole spectrum, not just a finite handful of eigenvalues, shifts under such a perturbation.

For our purposes, it is more useful to phrase this as an elimination statement for a linear system. If $(D-\lambda)\,x=UW^Tx+\text{source}$ with $D-\lambda$ invertible and $U,W$ of rank $r\ll\dim D$, one may solve for $x$ in terms of $(D-\lambda)^{-1}$ and $r$ scalar unknowns $W^Tx$, closing the system into an $r\times r$ linear problem. This is the content of the Feshbach/L\"owdin partitioning used in Sec.~\ref{sec:KKelimination}: the \emph{bulk} (i.e., the full KK tower) is solved for algebraically and fed back into a finite number of equations for the collective coordinates, i.e., the two scalars $\langle f|x\rangle,\langle k|x\rangle$ built from Eq.~\eqref{eq:fkdef}.

%======================================================================
\section{The discontinuity}
\label{app:sine}
%======================================================================
\noindent
Here we prove the analytic fact needed in Sec.~\ref{sec:sums} to evaluate $P$ in the order $N\to\infty$ first, $\epsilon\to0^+$ second. For real $a$ (not a negative integer) and $0<\theta<2\pi$, let $S(\theta,a)\equiv\sum_{n=1}^\infty\sin(n\theta)/(n+a)$, a conditionally convergent series by Dirichlet's test. Then
\begin{equation}
    \lim_{\theta\to0^+}S(\theta,a)=\frac\pi2,\label{eq:sinelimit}
\end{equation}
independently of $a$. To see this, note that the series
\begin{align*}
    \sum_{n=1}^\infty\Big[\dfrac{\sin(n\theta)}{n+a}-\dfrac{\sin(n\theta)}n\Big]=-a\sum_{n=1}^\infty\dfrac{\sin(n\theta)}{n(n+a)}
\end{align*}
converges absolutely and \emph{uniformly} in $\theta$, since its terms are bounded by the $\theta$-independent summable sequence $1/[n(n+a)]$ (for $n$ large enough that $n+a>0$; finitely many exceptional terms, if any, are continuous in $\theta$ and do not affect the limit). Dominated convergence therefore allows us to take $\theta\to0^+$ term by term, and each term $\sin(n\theta)/[n(n+a)]\to0$, so
\begin{equation*}
    \lim_{\theta\to0^+}\Big[S(\theta,a)-S(\theta,0)\Big]=0.
\end{equation*}
It remains to evaluate $S(\theta,0)=\sum_{n=1}^\infty\sin(n\theta)/n$, the classical Fourier series of the $2\pi$-periodic sawtooth wave,
\begin{equation*}
    S(\theta,0)=\operatorname{Im}\Big[-\ln(1-e^{i\theta})\Big]=\frac{\pi-\theta}2,\qquad 0<\theta<2\pi,
\end{equation*}
so that $\lim_{\theta\to0^+}S(\theta,0)=\pi/2$, and hence $\lim_{\theta\to0^+}S(\theta,a)=\pi/2$ for every $a$, proving Eq.~\eqref{eq:sinelimit}. The argument $a$ plays no special role here, which is why the limit in Sec.~\ref{sec:sums} comes out the same for $a=x$ and $a=-x$.

The same limit can be seen directly, without reference to the sawtooth series, by rescaling $u=n\theta$ and treating $\theta$ as the spacing of a Riemann sum:
\begin{equation}
S(\theta,a)=\sum_{n=1}^\infty\frac{\sin(n\theta)}{n+a}
=\sum_{n=1}^\infty\theta\,\frac{\sin(u_n)}{u_n+a\theta}
\ \xrightarrow{\ \theta\to0^+\ }\ \int_0^\infty\frac{\sin u}{u}\,du=\frac\pi2,
\end{equation}
since $a\theta\to0$ term by term and the sum converges to the Riemann integral in the variable $u$. Each term shrinks as $\theta\to0$, but so does the spacing between successive terms, and the two effects cancel exactly: the discrete sum, in this rescaled variable, becomes the classical Dirichlet integral, and inherits its value regardless of how small $\theta$ is. This is the same reason a truncated Fourier series overshoots a jump discontinuity by a fixed, universal amount however finely it is resolved.

This route also avoids any intermediate step through digamma or Lerch-transcendent asymptotics, as in Ref.~\cite{Barcelo:2014kha}'s discussion after their Eq.~(39), which would carry both a real, $a$-dependent piece and a term growing like $\ln\epsilon$, together with the Euler-Mascheroni constant $\gamma$. The final result, Eq.~\eqref{eq:noncommute}, comes out free of $\gamma$ and $\ln\epsilon$, matching the explicit remark in footnote~6 of Ref.~\cite{Barcelo:2014kha}.

%%%%%%%%%%%%%%%%%%%%%%%%%%%%%%%%%%%%%%%%%%%%%%%%%%%%%%
\begin{figure*}[!t]
    \captionsetup[subfigure]{labelformat=empty}
    \centering
    \subfloat%[{\bf a}]
    {\includegraphics[width=0.48\textwidth]{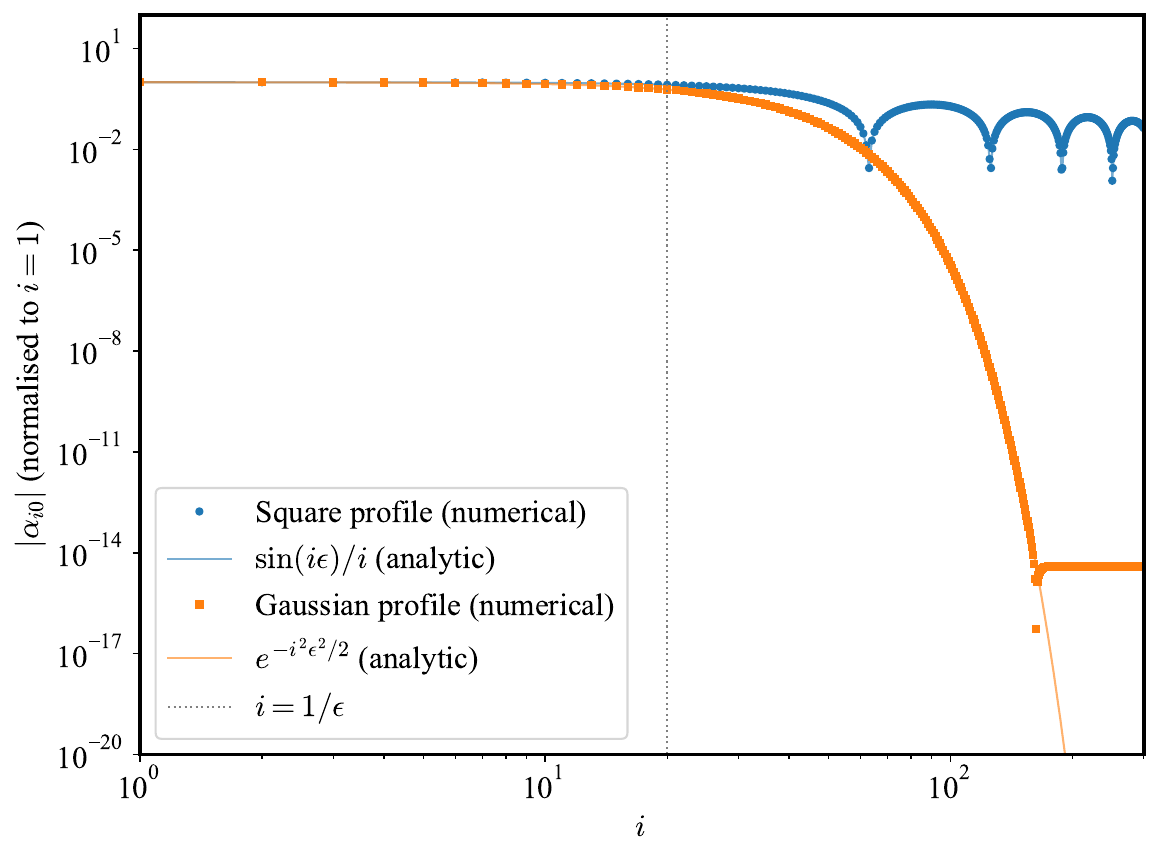}\label{fig:alpha_decay}}\hfill
    \subfloat%[{\bf b}]
    {\includegraphics[width=0.48\textwidth]{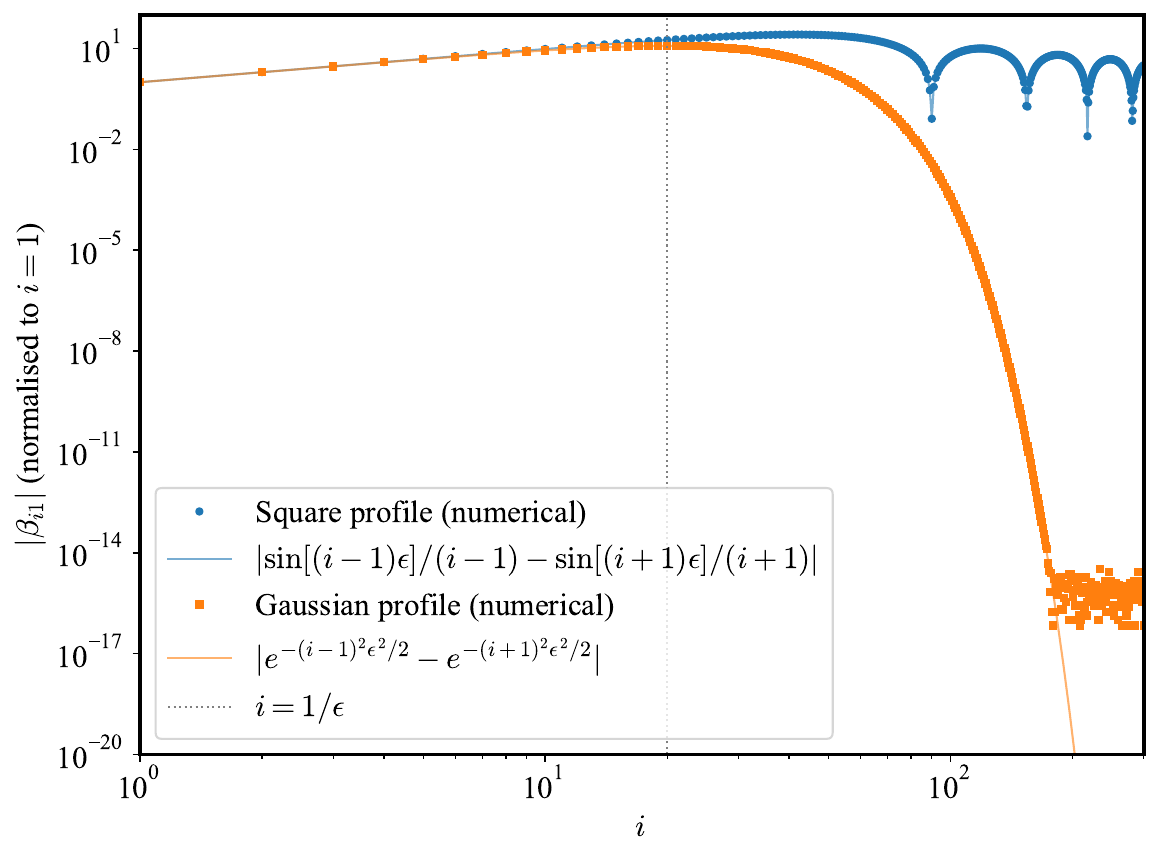}\label{fig:beta_decay}}
    \caption{Decay of the Yukawa matrix elements $|\alpha_{i0}|$ (left) and $|\beta_{i1}|$ (right) with KK level, for the smoothed square and Gaussian profiles at $\epsilon=0.05$: numerical integration (points) against the closed-form derived in Sec.~\ref{sec:sums} (solid lines), both normalised to $i=1$. The square profile's matrix elements persist, oscillating around a slowly falling envelope, well past $i=1/\epsilon$ (dotted line); the Gaussian's remain comparable there but fall away rapidly beyond it, reaching the double-precision noise floor by $i\sim10/\epsilon$.\label{fig:decay}}
\end{figure*}
%%%%%%%%%%%%%%%%%%%%%%%%%%%%%%%%%%%%%%%%%%%%%%%%%%%%%%
%======================================================================
\section{Numerical check of the profile decay rates}
\label{app:numerics}
%======================================================================
\noindent
In Sec.~\ref{sec:sums}, we argue that the square profile's matrix elements decay only as a power law in the KK-level combination $n=i\pm j$, while a profile vanishing smoothly at the edges of its support -- a Gaussian, in particular -- decays faster than any power of $n$. We validate both claims numerically by computing the overlap integrals.

Setting $R=1$ without loss of generality, we first replace the boundary regulator by a bump $\rho_\epsilon(y)$ of width $\epsilon$, i.e., take $\rho_\epsilon(y)=1/\epsilon$ for $y\in[\pi-\epsilon,\pi]$, and consider a Gaussian profile $\rho_\epsilon(y)\propto\exp[-(y-\pi)^2/2\epsilon^2]$, both normalised to integrate to one. We then evaluate the lowest nonzero off-diagonal matrix elements
\begin{equation}
    \alpha_{i0}=\int_0^\pi\!dy\,\rho_\epsilon(y)\,q_L^i(y)\,q_L^0(y),\qquad
\beta_{i1}=\int_0^\pi\!dy\,\rho_\epsilon(y)\,q_R^i(y)\,q_R^1(y),
\end{equation}
by direct numerical quadrature on a fine grid. Figure~\ref{fig:decay} shows the result for $\epsilon=0.05$ from $i=1$ to $i=300$. Both agree with our argument in the main text. The square profile's matrix elements are still $O(1/10)$ of their $i=1$ value at $i=5/\epsilon$, oscillating rather than falling monotonically; the Gaussian's remain comparable to the square profile's through $i\sim1/\epsilon$, but fall roughly five orders of magnitude below them by $i=5/\epsilon$, and reach the double-precision noise floor only by $i\sim10/\epsilon$. This confirms numerically that the obstruction discussed in Sec.~\ref{sec:sums} is a property of the square profile's sharp edges and not of smoothing the Higgs peak in general.

\bibliography{references}

@article{Gopalakrishna:2013hua,
    author = "Gopalakrishna, Shrihari and Mandal, Tanumoy and Mitra, Subhadip and Moreau, Gr{\'e}gory",
    title = "{LHC Signatures of Warped-space Vectorlike Quarks}",
    eprint = "1306.2656",
    archivePrefix = "arXiv",
    primaryClass = "hep-ph",
    doi = "10.1007/JHEP08(2014)079",
    journal = "JHEP",
    volume = "08",
    pages = "079",
    year = "2014"
}

@article{Barcelo:2014kha,
    author = "Barcel{\'o}, Roberto and Mitra, Subhadip and Moreau, Gr{\'e}gory",
    title = "{On a boundary-localized Higgs boson in 5D theories}",
    eprint = "1408.1852",
    archivePrefix = "arXiv",
    primaryClass = "hep-ph",
    reportNumber = "LPT-ORSAY-14-68, LPT-Orsay-14-68",
    doi = "10.1140/epjc/s10052-015-3756-3",
    journal = "Eur. Phys. J. C",
    volume = "75",
    number = "11",
    pages = "527",
    year = "2015"
}

@article{Randall:1999ee,
    author = "Randall, Lisa and Sundrum, Raman",
    title = "{A Large mass hierarchy from a small extra dimension}",
    eprint = "hep-ph/9905221",
    archivePrefix = "arXiv",
    reportNumber = "MIT-CTP-2860, PUPT-1860, BUHEP-99-9",
    doi = "10.1103/PhysRevLett.83.3370",
    journal = "Phys. Rev. Lett.",
    volume = "83",
    pages = "3370-3373",
    year = "1999"
}

@article{Gherghetta:2000qt,
    author = "Gherghetta, Tony and Pomarol, Alex",
    title = "{Bulk fields and supersymmetry in a slice of AdS}",
    eprint = "hep-ph/0003129",
    archivePrefix = "arXiv",
    reportNumber = "CERN-TH-2000-081, UNIL-IPT-00-06",
    doi = "10.1016/S0550-3213(00)00392-8",
    journal = "Nucl. Phys. B",
    volume = "586",
    pages = "141-162",
    year = "2000"
}

@article{Csaki:2003sh,
    author = "Csaki, Csaba and Grojean, Christophe and Hubisz, Jay and Shirman, Yuri and Terning, John",
    title = "{Fermions on an interval: Quark and lepton masses without a Higgs}",
    eprint = "hep-ph/0310355",
    archivePrefix = "arXiv",
    reportNumber = "SACLAY-T03-147",
    doi = "10.1103/PhysRevD.70.015012",
    journal = "Phys. Rev. D",
    volume = "70",
    pages = "015012",
    year = "2004"
}

@article{Feshbach1958,
    author = "Feshbach, Herman",
    title = "{Unified theory of nuclear reactions}",
    journal = "Ann. Phys.",
    volume = "5",
    number = "4",
    pages = "357-390",
    year = "1958",
    doi = "10.1016/0003-4916(58)90007-1"
}

@article{Feshbach1962,
    author = "Feshbach, Herman",
    title = "{A unified theory of nuclear reactions. II}",
    journal = "Ann. Phys.",
    volume = "19",
    number = "2",
    pages = "287-313",
    year = "1962",
    doi = "10.1016/0003-4916(62)90221-X"
}

@article{Lowdin1951,
    author = {L\"owdin, Per-Olov},
    title = "{A Note on the Quantum-Mechanical Perturbation Theory}",
    journal = "J. Chem. Phys.",
    volume = "19",
    number = "11",
    pages = "1396-1401",
    year = "1951",
    doi = "10.1063/1.1748067"
}

@article{WA1,
    author = "Weinstein, A.",
    title = "{\'Etude des spectres des \'equations aux d\'eriv\'ees partielles de la th\'eorie des plaques \'elastiques}",
    journal = "M\'em. Soc. Math. France",
    volume = "88",
    pages = "1-56",
    year = "1937"
}

@article{WA2,
    author = "Aronszajn, N.",
    title = "{The Rayleigh-Ritz and A. Weinstein methods for approximation of eigenvalues, I-II}",
    journal = "Proc. Natl. Acad. Sci. U.S.A.",
    volume = "34",
    number = "10",
    pages = "474-480",
    year = "1948",
    doi = "10.1073/pnas.34.10.474"
}

@article{WA3,
    author = "Howland, James S.",
    title = "{On the Weinstein-Aronszajn formula}",
    journal = "Arch. Rational Mech. Anal. ",
    volume = "39",
    pages = "323-339",
    year = "1970",
    doi = "10.1007/BF00251295"
}

@article{Dienes:1998sb,
    author = "Dienes, Keith R. and Dudas, Emilian and Gherghetta, Tony",
    title = "{Neutrino oscillations without neutrino masses or heavy mass scales: A Higher dimensional seesaw mechanism}",
    eprint = "hep-ph/9811428",
    archivePrefix = "arXiv",
    reportNumber = "CERN-TH-98-370",
    doi = "10.1016/S0550-3213(99)00377-6",
    journal = "Nucl. Phys. B",
    volume = "557",
    pages = "25",
    year = "1999"
}

@article{Arkani-Hamed:1998wuz,
    author = "Arkani-Hamed, Nima and Dimopoulos, Savas and Dvali, G. R. and March-Russell, John",
    title = "{Neutrino masses from large extra dimensions}",
    eprint = "hep-ph/9811448",
    archivePrefix = "arXiv",
    reportNumber = "SLAC-PUB-8014, SU-ITP-98-64",
    doi = "10.1103/PhysRevD.65.024032",
    journal = "Phys. Rev. D",
    volume = "65",
    pages = "024032",
    year = "2001"
}

@article{Choi:2015fiu,
    author = "Choi, Kiwoon and Im, Sang Hui",
    title = "{Realizing the relaxion from multiple axions and its UV completion with high scale supersymmetry}",
    eprint = "1511.00132",
    archivePrefix = "arXiv",
    primaryClass = "hep-ph",
    doi = "10.1007/JHEP01(2016)149",
    journal = "JHEP",
    volume = "01",
    pages = "149",
    year = "2016"
}

@article{Giudice:2016yja,
    author = "Giudice, Gian F. and McCullough, Matthew",
    title = "{A Clockwork Theory}",
    eprint = "1610.07962",
    archivePrefix = "arXiv",
    primaryClass = "hep-ph",
    doi = "10.1007/JHEP02(2017)036",
    journal = "JHEP",
    volume = "02",
    pages = "036",
    year = "2017"
}

@article{Sherman:1950,
    author = "Sherman, Jack and Morrison, Winifred J.",
    title = "{Adjustment of an Inverse Matrix Corresponding to a Change in One Element of a Given Matrix}",
    journal = "Ann. Math. Statist.",
    volume = "21",
    number = "1",
    pages = "124-127",
    year = "1950",
    doi = "10.1214/aoms/1177729893"
}

@techreport{Woodbury:1950,
    author = "Woodbury, Max A.",
    title = "{Inverting Modified Matrices}",
    institution = "Statistical Research Group, Princeton University",
    number = "Memorandum Report 42",
    address = "Princeton, NJ",
    year = "1950"
}

@article{Krein:1953,
    author = "Krein, Mark G.",
    title = "{On the Trace Formula in Perturbation Theory}",
    journal = "Mat. Sbornik N.S.",
    volume = "33(75)",
    pages = "597-626",
    year = "1953",
    note = "In Russian"
}

@article{Malm:2013jia,
    author = "Malm, Raoul and Neubert, Matthias and Novotny, Kristiane and Schmell, Christoph",
    title = "{5D Perspective on Higgs Production at the Boundary of a Warped Extra Dimension}",
    eprint = "1303.5702",
    archivePrefix = "arXiv",
    primaryClass = "hep-ph",
    reportNumber = "MITP-13-01",
    doi = "10.1007/JHEP01(2014)173",
    journal = "JHEP",
    volume = "01",
    pages = "173",
    year = "2014"
}

@article{Carena:2012fk,
    author = "Carena, Marcela and Casagrande, Sandro and Goertz, Florian and Haisch, Ulrich and Neubert, Matthias",
    title = "{Higgs Production in a Warped Extra Dimension}",
    eprint = "1204.0008",
    archivePrefix = "arXiv",
    primaryClass = "hep-ph",
    reportNumber = "MZ-TH-12-13, FERMILAB-PUB-12-884-T",
    doi = "10.1007/JHEP08(2012)156",
    journal = "JHEP",
    volume = "08",
    pages = "156",
    year = "2012"
}

@article{Azatov:2009na,
    author = "Azatov, Aleksandr and Toharia, Manuel and Zhu, Lijun",
    title = "{Higgs Mediated FCNC's in Warped Extra Dimensions}",
    eprint = "0906.1990",
    archivePrefix = "arXiv",
    primaryClass = "hep-ph",
    reportNumber = "UMD-PP-09-039",
    doi = "10.1103/PhysRevD.80.035016",
    journal = "Phys. Rev. D",
    volume = "80",
    pages = "035016",
    year = "2009"
}

@article{Azatov:2010pf,
    author = "Azatov, Aleksandr and Toharia, Manuel and Zhu, Lijun",
    title = "{Higgs Production from Gluon Fusion in Warped Extra Dimensions}",
    eprint = "1006.5939",
    archivePrefix = "arXiv",
    primaryClass = "hep-ph",
    reportNumber = "UMD-PP-10-010",
    doi = "10.1103/PhysRevD.82.056004",
    journal = "Phys. Rev. D",
    volume = "82",
    pages = "056004",
    year = "2010"
}

@article{Casagrande:2010si,
    author = "Casagrande, Sandro and Goertz, Florian and Haisch, Uli and Neubert, Matthias and Pfoh, Torsten",
    title = "{The Custodial Randall-Sundrum Model: From Precision Tests to Higgs Physics}",
    eprint = "1005.4315",
    archivePrefix = "arXiv",
    primaryClass = "hep-ph",
    reportNumber = "MZ-TH-10-18",
    doi = "10.1007/JHEP09(2010)014",
    journal = "JHEP",
    volume = "09",
    pages = "014",
    year = "2010"
}

\end{document}